\documentclass[MSNbibl,nameyear,dvips]{arxstspdf}
\usepackage{flushend}
\usepackage{stfloats}
\volume{29}
\issue{2}
\pubyear{2014}
\firstpage{259}
\lastpage{260}
\doi{10.1214/14-STS475} % kopijuoti is 'New paper accepted'
\referstodoi{10.1214/13-STS457}% pagrindinio straipsnio DOI, kai straipsnis yra diskusija ar rejoinder'is



\begin{document}
\begin{frontmatter}

\vspace*{6pt}\title{Discussion of ``On the Birnbaum Argument for the Strong Likelihood
Principle''}%\thanksref{T1}
% kai straipsnis turi susijusiu diskusiju ir rejoinder'iu
\runtitle{Discussion}

\begin{aug}
\author[a]{\fnms{Jan F.}~\snm{Bj\o rnstad}\corref{}\ead[label=e1]{jab@ssb.no}}
\runauthor{J. F. Bj\o rnstad}

\affiliation{Statistics Norway}

\address[a]{Jan F. Bj\o rnstad is Professor of Statistics, Department of Mathematics,
University of Oslo and Head of Research, Division for
Statistical Methods, Statistics Norway,
P.O. Box 8131 Dep., N-0033 Oslo,
Norway \printead{e1}.}
\end{aug}

% ABSTRACT
\begin{abstract}
The paper by Mayo claims to provide a new clarification
and critique of Birnbaum's argument for showing that sufficiency and
conditionality principles imply the likelihood principle. However, much of
the arguments go back to arguments made thirty to forty years ago. Also,
the main contention in the paper, that Birnbaum's arguments are not valid,
seems to rest on a misunderstanding.
\end{abstract}

% KEYWORDS
% Pirmas kwd is didziosios raides
\begin{keyword}
\kwd{Likelihood}
\kwd{conditionality}
\kwd{sufficiency}
\kwd{Birnbaum's theorem}
\end{keyword}
\end{frontmatter}

The goal of this paper is to provide a new clarification and critique of
Birnbaum's argument for showing that principles of sufficiency and
conditionality entail the (strong) likelihood principle (LP).

I must admit I do not find that the paper provides such a new clarification
of the criticism of Birnbaum's argument. Rather, much of the criticism in
the paper goes back to arguments made in the 70s and 80s by several
authors, for example, Durbin (\citeyear{6}), \citet{8}, \citet{4} and
\citet{7}. This critique has been discussed by several
statisticians with an opposing view; see \citet{1} and Bj\o rnstad (\citeyear{2}).

I will concentrate my discussion on what seems to be the most important
contention in the paper, that the sufficient statistic in Birnbaum's proof
erases the information as to which experiment the data came from and,
hence,\vadjust{\goodbreak} that the weak conditionality principle (WCP) cannot be applied; see,
for example, Sections~5.2 and~7.

As I understand it, this is a misunderstanding of the proof. For one thing,
it seems that only the observation $x_{2}$ from the experiment $E_{2}$ is
considered in the mixture experiment instead of the correct ($E_{2}$,
$x_{2}$). The observations in a mixture experiment are always of the form
($E_{h}$, $x_{h})$---\textit{never} as only $x_{h}$.

Other arguments leading up to this contention seem to rest on a
misunderstanding of the sufficiency considerations in the proof, that given
an observation from a certain experiment the result in an unperformed
experiment is to be reported; see, for example, Sections~2.4 and 5.1. I
find that this is definitely not the case. To be specific, the author
considers the following proof in the discrete case:

Let ($j,x_{j}$) indicate that Experiment $E_{j}$ was performed with data
$x_{j}, j =1, 2$. Assume the data values $x_{1}^{0}$ and $x_{2}^{0}$ have
proportional likelihoods from experiments $E_{1}$ and $E_{2}$, respectively.
Then the sufficient statistic in the mixture experiment used in Birnbaum's
proof is given by
\begin{eqnarray*}
T(j, x_{j})& =& (j, x_{j}) \quad \mbox{if } (j,
x_{j}) \neq \bigl(1, x_{1}^{0}\bigr), \bigl(2,
x_{2}^{0}\bigr),
\\
T\bigl(1, x_{1}^{0}\bigr)& =& T\bigl(2,
x_{2}^{0}\bigr) = \bigl(1, x_{1}^{0}
\bigr).
\end{eqnarray*}
Mayo claims that $T(1, x_{1}^{0}) = T(2, x_{2}^{0})\ (= c)$ implies that
the weak conditionality principle (WCP) is violated. Now, one should note
that the proof works with any $c \neq (1, x_{1}^{0})$ and
$c \neq (j,x_{j}), j = 1, 2$ and all $x_{j}$. When $E_{2}$ is\vadjust{\goodbreak}
performed and $x_{2}^{0}$ is the result, then the evidence should only
depend on $E_{2}$ and $x_{2}^{0}$, and not on a result of an unperformed
experiment $E_{1}$. This is, of course, correct, but it does not depend on
a result in $E_{1}$. (Actually $E_{1}$ is not an unperformed experiment
either. We comment on this issue later.) By letting $c = (1,
x_{1}^{0})$ it seems so, but we see by choosing a $c \neq (1,
x_{1}^{0})$ it is not the case. So WCP is not violated. The sufficient
statistic simply takes the same value for these two results of the mixture
experiment. It has nothing to do with WCP.

So Birnbaum's proof does not require that the evidential support of a known
result should depend on the result of an unperformed experiment. It follows
that the main contention in the paper seems to rest on a lack of
understanding of the basics of the proof of Birnbaum's theorem. In fact, it
is possible to do the proof even more generally. One can show, for a given
experiment [see, e.g., \citet{5} and Bj\o rnstad (\citeyear{3})], that
if two likelihoods are proportional for two possible observations in the
same experiment, there exists a minimal sufficient statistic with the same
value for the two observations. This holds both for discrete and continuous
models.

To make the sufficiency argument clearer, consider a mixture of a binomial
experiment $E_{1}$ and a negative binomial experiment $E_{2}$ where the
observations are $x_{1} =$ number of successes in 12 trials and $x_{2} =$
number of trials until 3 successes. If $x_{1}^{0} = 3$ and $x_{2}^{0} x_{2}
= 12$ then the likelihoods are proportional. A natural choice of the
sufficient statistic $T$ in the mixture experiment in Birnbaum's proof
has
\begin{eqnarray*}
&&T\bigl(1, x_{1}^{0}\bigr) = T\bigl(2, x_{2}^{0}
\bigr) = 3/12,\\
&&\quad \mbox{the proportion of successes in either case}.
\end{eqnarray*}
Clearly then, the value of $T$ from experiment $E_{2}$ does not depend on
the result from $E_{1}$.

As already mentioned, the author claims that the sufficient statistic $T$
in the proof of Birnbaum's result has the effect of erasing the index of
the experiment. Moreover, it is claimed that inference based on $T$ is to
be computed over the performed and unperformed experiments $E_{1}$ and
$E_{2}$. As we have shown, this is simply not the case. It should also be
mentioned that statistically the proof simply considers two instances of
performing the mixture experiments resulting in proportional likelihoods
and really has nothing to do with considering unperformed experiments.

Let me also mention that the author's premise in Section~5 is not correct.
The starting point is \textit{not} that we have an arbitrary outcome of one
single experiment, but rather that two experiments have been performed
about the same parameter resulting in proportional likelihoods. So Birnbaum
does not enlarge a known single experiment but constructs a mixture of the
two performed experiments. There is really $no$ \textit{unperformed}
experiment here. In a sense, one may regard the paper by Mayo as actually
not discussing the LP at all.

It should be clear that I find that the main contention in the paper does
not hold when maintaining the original meaning of the principles of
sufficiency, conditionality and likelihood. Other comments made in this
paper referring to various authors in the 70s and 80s are a different
matter. However, I do not find any \textit{new} clarification of Birnbaum's
fundamental theorem in this paper. For example, regarding sufficiency, it
is necessary to restrict the application of sufficiency to nonmixture
experiments, as \citet{8} did, in order to invalidate Birnbaum's
result. \citet{1} argue, I think, convincingly against such
a restriction. See also Bj\o rnstad (\citeyear{2}).

Let me end this discussion by making clear the following fact: It is
obviously clear that frequentistic measures may, and typically do, violate
LP. This is true as far as it comes to analysis of the actual data we
observe. But a major point here is that the LP does not say that one should
not be concerned with how the methods do when used repeatedly. LP is simply
\textit{not about method evaluation}. Evaluation of methods is still
important. So LP says in essence that frequentistic considerations are not
\textit{sufficient} for evaluating the uncertainty and reliability in the
statistical analysis of the actual data; see also Bj\o rnstad (\citeyear{3}) for a
discussion on this issue.

% zodis "Acknowledgments" paliekamas pagal autoriu

%suskaldyti doi

% imsref loaded by jurgita.kaciuliene, 2014-05-28 15:30:18

\end{document}